\documentclass{svproc}
\usepackage{url}
\usepackage{graphicx,color}
\usepackage{epsfig}
\usepackage{amssymb}
\usepackage[dvipsnames]{xcolor}

\begin{document}
\mainmatter              
\title{{\it Ab initio} folding potentials for proton-nucleus scattering with 
NCSM nonlocal one-body densities}
\titlerunning{{\it Ab initio} folding potentials}  
%
\author{Ch. Elster\inst{1} \and M. Burrows\inst{1}
S.P. Weppner\inst{2} \and K.D.~Launey\inst{3} \and P.~Maris\inst{4} \and G.~Popa\inst{1}} 
\authorrunning{Ch. Elster et al.} 
\institute{Dept. of Physics and Astronomy,  Ohio University, Athens, OH 45701,
USA,\\
\and
Natural Sciences, Eckerd College, St. Petersburg, FL 33711, USA \\
\and
Dept. of Physics and Astronomy, Louisiana State
University, Baton Rouge, LA 70803, USA \\
\and 
Dept. of Physics and Astronomy, Iowa State University,
Ames, IA 50011, USA}

\maketitle              

\begin{abstract}
Based on the spectator expansion of the multiple scattering series we employ a nonlocal
translationally invariant nuclear density derived from a chiral next-to-next-to-leading order
(NNLO) and the very same interaction for consistent full-folding calculations of the
effective (optical) potential for nucleon-nucleus scattering for light nuclei.
\end{abstract}

The theoretical approach to the elastic scattering of a nucleon from a
  nucleus, pioneered by Watson \cite{Watson1953a}, made familiar by
  Kerman, McManus, and Thaler (KMT) \cite{KMT} and further developed as
  the spectator expansion \cite{Siciliano:1977zz,Chinn:1995qn} is now being
applied together with {\it ab initio} structure calculations to obtain
effective folding nucleon-nucleus (NA) potentials. 
The spectator expansion is predicated upon the idea that at projectile energies
higher than about 100~MeV  the two-body interactions between 
the projectile and the nucleons in the target dominate elastic scattering, for
which a transition operator can be defined as
\begin{equation}
  T_{el}\equiv PTP= PUP +PUPG_0(E) T_{el} \label{eq:1}.
\end{equation} 
The projector $P=\frac{|\Phi_{A}\rangle\langle\Phi_{A}|}{\langle\Phi_{A}|\Phi_{A}
\rangle}$ is conventionally taken to project on the elastic channel so that
$[G_0,P]=0$. Here $|\Phi_{A}\rangle$ stands for the ground state of the target so that
$H_A |\Phi_{A}\rangle =E_A |\Phi_{A}\rangle$, and $G_0(E)=(E-H_0+i\varepsilon)^{-1}$,
where $H_0=h_0+H_A$, being the  free propagator for the projectile+target
system. The effective (optical) potential is given by 
\begin{equation}
U = V + V G_0(E) Q U, \label{eq:2}
\end{equation}
where the operator $Q$ is defined via the relation $P+Q=1$. The first-order term
involves two-body interactions between
the projectile and one of the target nucleons, {\it i.e.} $U =
\sum_{i=1}^{A}\tau_{i}$, where the operator $\tau_{i}$ is derived to be
\begin{eqnarray}
\tau_i &=& v_{0i} + v_{0i} G_0(E) Q \tau_i \nonumber \\
&=& v_{0i} + v_{0i}G_0(E) \tau_i - v_{0i} G_0(E) P \tau_i
\label{eq:3} \\
&=& \hat{\tau}_i - \hat{\tau}_i G_0(E) P \tau_i . \nonumber
\end{eqnarray}
Here $\hat{\tau}_i$ is the NN t-matrix and is defined as the solution of
$\hat{\tau}_i = v_{0i} + v_{0i} G_0(E) \hat{\tau}_i$. 
In lowest order $\hat{\tau}_i\approx t_{0i}$, which corresponds to the conventional
impulse approximation. Here the operator $t_{0i}$ stands for the standard solution of a
Lippmann-Schwinger equation with the NN interaction as driving term.

For elastic scattering only $P\tau_i P$ 
 needs to be considered, 
\begin{equation}
\langle\Phi_A| \tau_i | \Phi_A\rangle = \langle\Phi_A| \hat{\tau_i}|
 \Phi_A\rangle - \langle\Phi_A| \hat{\tau_i}| \Phi_A\rangle \frac {1}
 {(E-E_A) - h_0 + i\varepsilon} \langle\Phi_A| \tau_i | \Phi_A\rangle ,
 \label{eq:4}
\end{equation}
and this matrix element determines the full-folding effective (optical)
potential when summing over all target nuclei,
\begin{equation}
\langle{\bf k}' | U |{\bf k}\rangle =
\langle{\bf k}'\Phi_{A} | \sum_{i} {\tau_{i}}
|{\bf k}\Phi_{A}\rangle . \label{eq:5}
\end{equation}
Since $\langle{\bf k}' | U |{\bf k}\rangle$ is the solution of the sum
of one-body integral equations represented by Eq.~(\ref{eq:4}),
it is sufficient to consider the driving term
\begin{equation}
\langle{\bf k}' |\hat{U}|{\bf k}\rangle =
\langle{\bf k}'\Phi_{A} | \sum_{i} \hat{\tau}_{i}
|{\bf k}\Phi_{A}\rangle , \label{eq:6}
\end{equation}
where $\hat{\tau}_{i} \approx t_{0i}$  when considering the first order single scattering term. 
\begin{figure}
{\epsfxsize=6.1cm
\epsffile{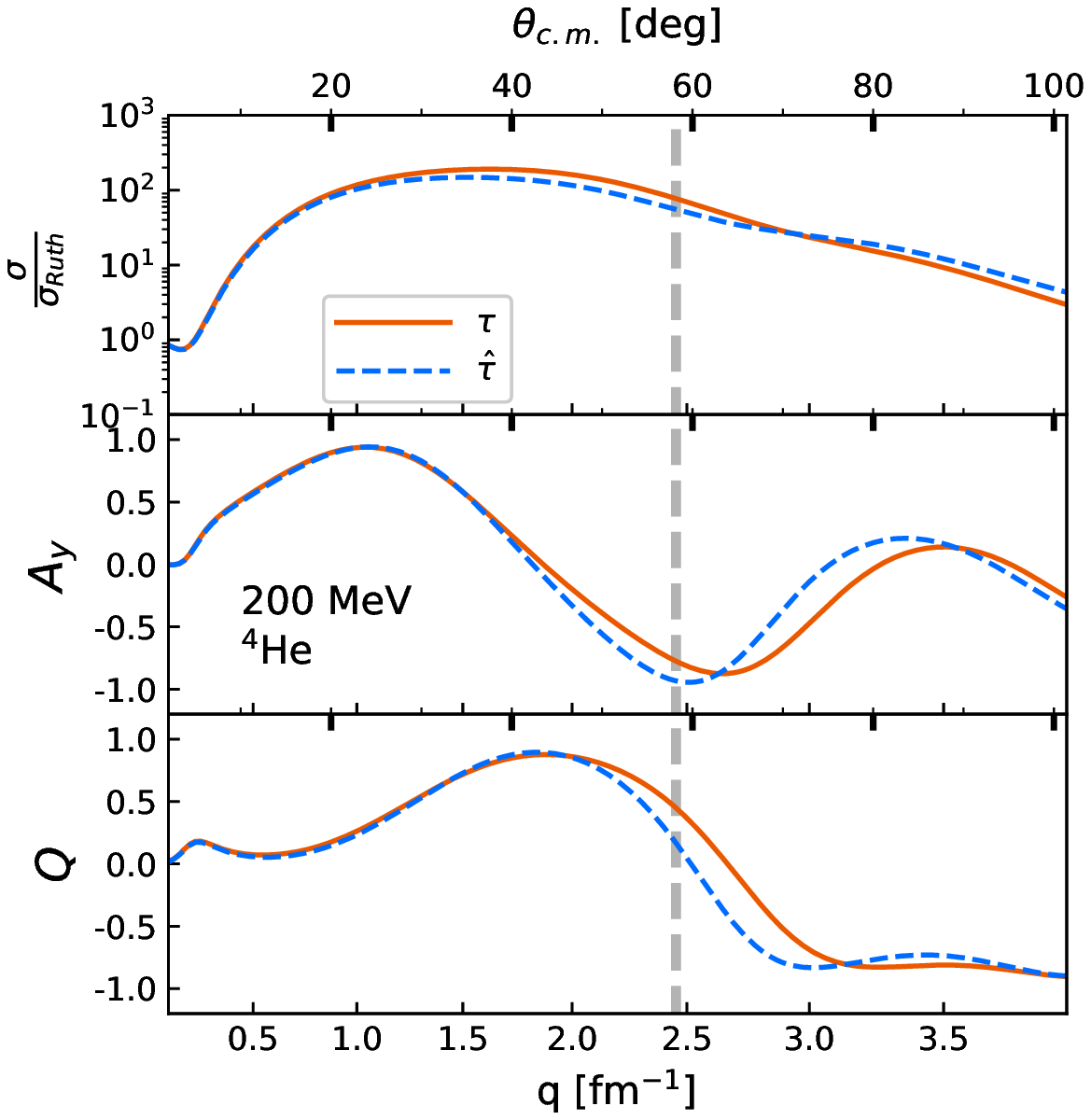}} 
\hspace{-3mm}
{\epsfxsize=6.1cm
\epsffile{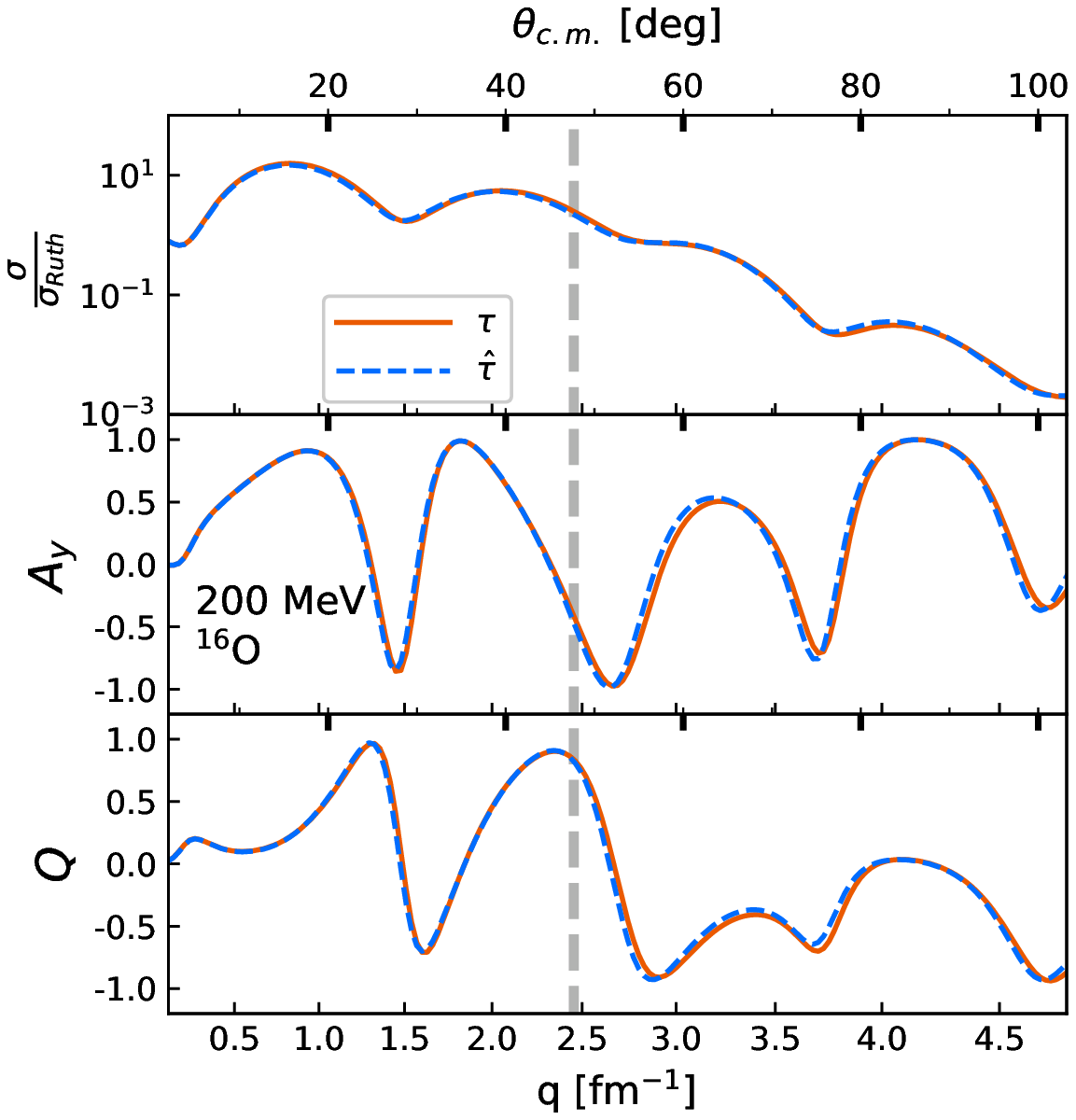}}
\caption{Angular distribution of the differential cross section divided by the
Rutherford cross section, analyzing power $A_y$ and spin rotation function $Q$ for
elastic proton scattering from $^4$He (left) and $^{16}$O (right) as
function of the momentum transfer and the c.m. angle calculated with the NNLO$_{\rm opt}$
chiral interaction~\protect\cite{Ekstrom13}. The solid line represents the calculation
based on $\tau_i$ (Eq.~(\ref{eq:5})), the dashed line the one based on ${\hat \tau}_i$
(Eq.~(\ref{eq:6})).
\label{fig1}
}
\end{figure}
Inserting a complete set of momenta for the struck target nucleon before and after the collision,
representing the sum over target protons and neutrons by $\alpha$, evaluating the momentum
conserving delta function and changing variables to ${\bf q}={\bf k'}-{\bf k}$, ${\bf
K}=\frac{1}{2}({{\bf k} + {\bf k^{\prime}}})$ and 
${\bf P}=\frac{1}{2}({\bf p}'+{\bf p}) + \frac{{\bf K}}{A}$, 
the final expression for the full-folding effective potential is given by
\begin{eqnarray}
\hat{U}({\bf q},{\bf K})=\sum_{\alpha=p,n}
\int d^{3}{\bf P}\;
\eta({\bf P},{\bf q},{\bf K})\;
\hat{\tau}_{\alpha}\left({\bf q},\frac{1}{2}\left(\frac{A+1}{A}{\bf K}-{\bf P}\right);
\epsilon\right)\; \nonumber \\
\;\;\;\;\;\;\;\;\;\;\;\;\;\;\;
\times \; \rho_{\alpha}\left({\bf P}-\frac{A-1}{A}\frac{\bf q}{2},
{\bf P}+\frac{A-1}{A}\frac{\bf q}{2}\right). \label{eq:7}
\end{eqnarray}
Here $\eta({\bf P},{\bf q},{\bf K})$ is the M\o ller factor for the frame
transformation~\cite{Joachain} relating the NN zero-momentum frame to the NA zero-momentum frame.
Further details can be found in~\cite{Burrows:2018ggt}. The quantity $\rho_{\alpha}$, with $\alpha
= p(n)$, represents a nonlocal one-body density matrix (OBD) for the proton (neutron) distribution,
and must be given in a translationally invariant fashion~\cite{Burrows:2017wqn}.
\begin{figure}
{\epsfxsize=6.1cm
\epsffile{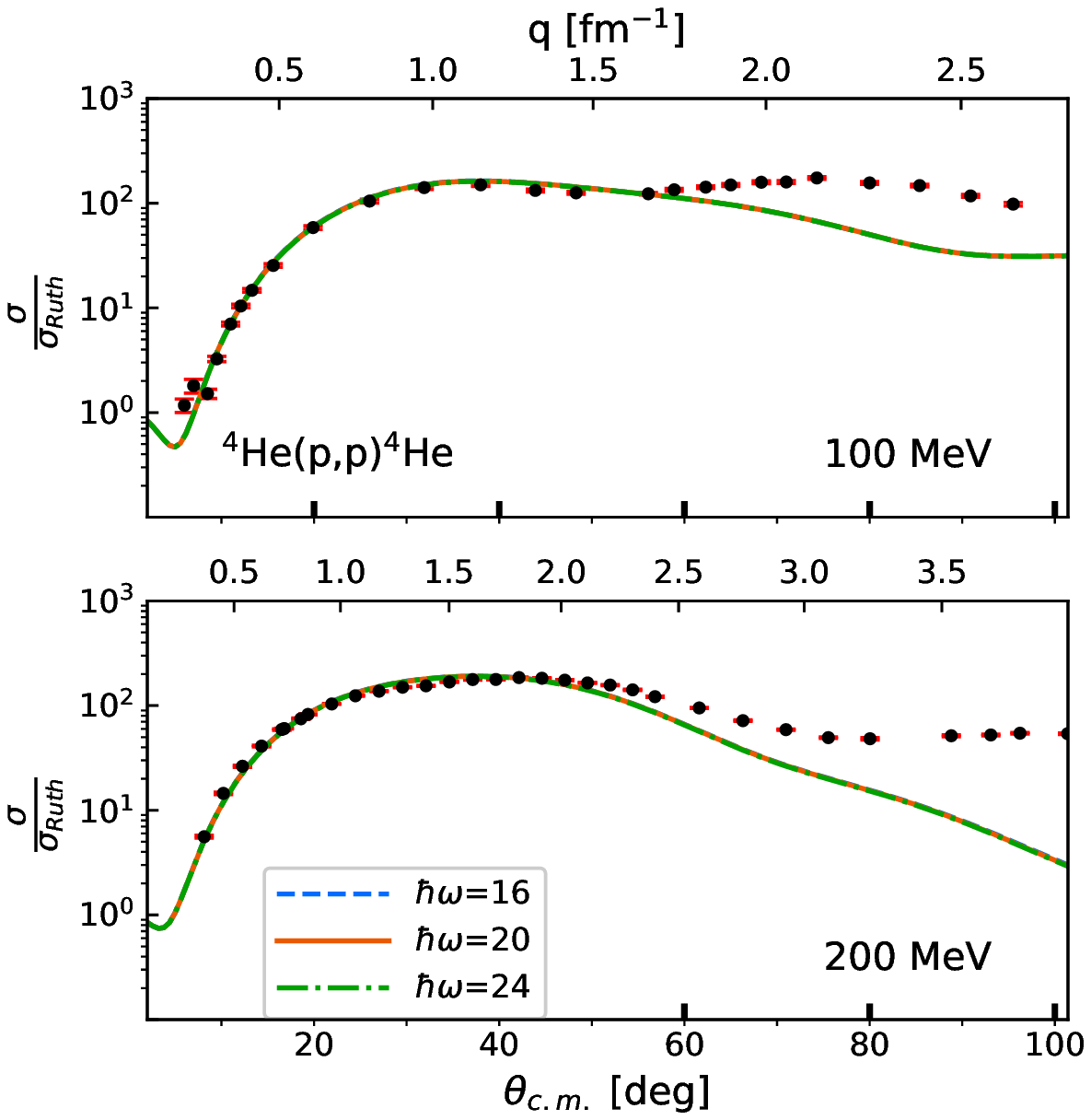}} 
\hspace{-3mm}
{\epsfxsize=6.1cm
\epsffile{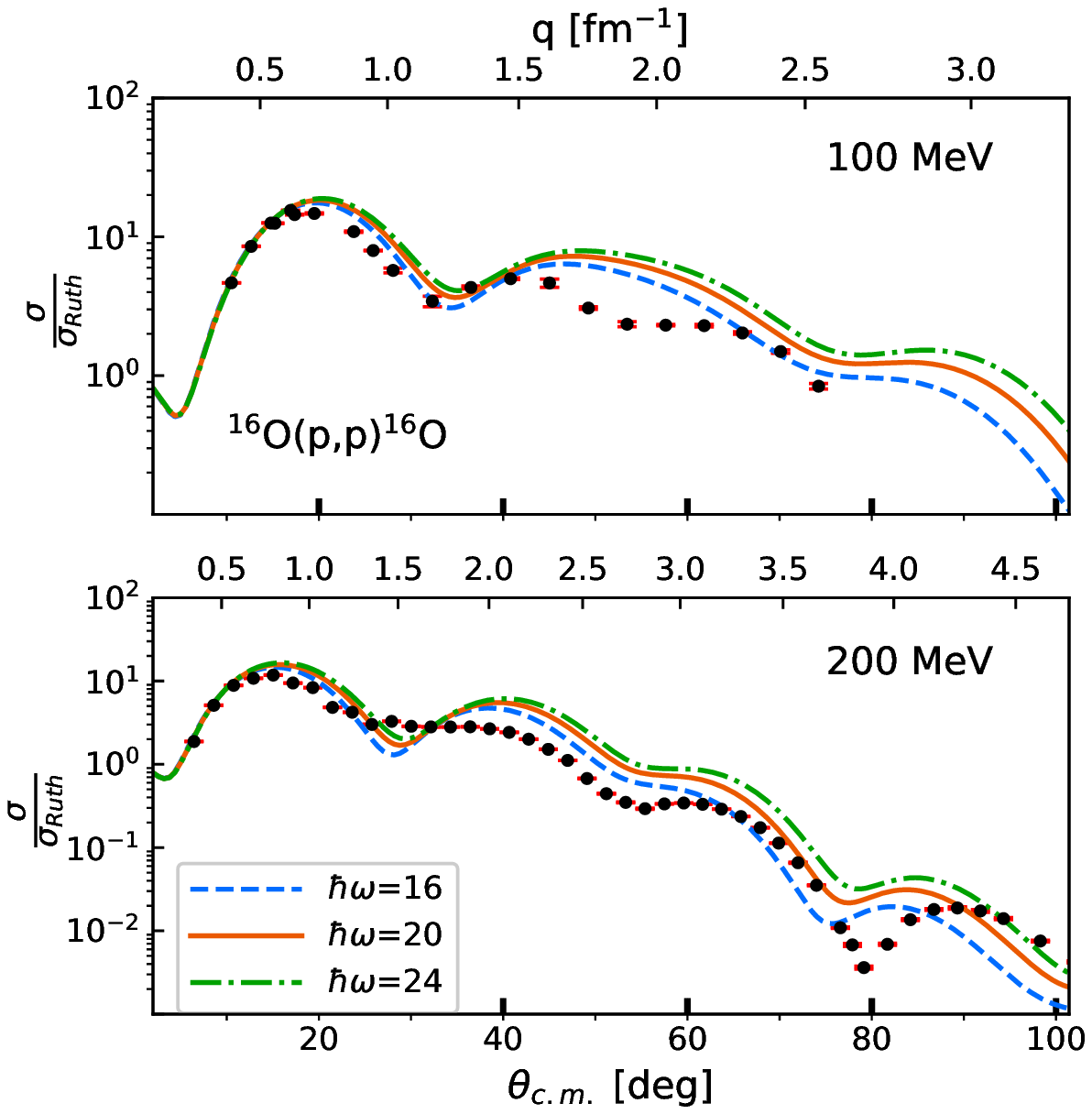}}
\caption{Angular distribution of the differential cross section divided by the
Rutherford cross section for
elastic proton scattering from $^4$He (left) and $^{16}$O (right) as
function of the c.m. angle calculated with the NNLO$_{\rm opt}$
chiral interaction~\protect\cite{Ekstrom13}. The calculations for $^4$He
are carried out with $N_{\rm max}=$~18, while the ones for $^{16}$O with
$N_{\rm max}=$~10. The values of $\hbar\omega$ are indicated in the
lower panels. The $^4$He data for 100~MeV are taken 
from~\cite{Goldstein:1970dg} and for 200~MeV from~\cite{Moss:1979aw}. The $^{16}$O data for 100~MeV are taken 
from~\cite{Seifert:1990um} and for 200~MeV from~\cite{Glover:1985xd}.
}
\label{fig2}
\end{figure}
An important product of this work is that the NN t-matrix and OBD now use the same underlying NN interaction.
For this we choose the optimized chiral NN interaction at 
next-to-next-to-leading order NNLO$_{\rm{opt}}$ from Ref.~\cite{Ekstrom13}. In the $A$~=~3, 4 nucleon systems the contributions of the 3NFs are smaller than in most other 
parameterizations of chiral interactions. From this point of view, the NNLO$_{\rm{opt}}$
interaction is very well suited for our calculations, since the first-order folding potential does not contain any explicit 3NF
contributions.
We calculated the full-folding integral for the first-order effective
(optical) potential for NA scattering  {\it ab initio}. i.e. they are
based consistently on one single NN interaction, in our case the chiral
  NNLO$_{\rm{opt}}$ interaction from Ref.~\cite{Ekstrom13}, which is
fitted to NN data up to 125~MeV laboratory kinetic energy
  with $\chi^2 \approx 1$ per degree of freedom. Based on this
interaction the one-body nonlocal, translationally invariant
 nuclear densities are calculated as laid
out in Ref.~\cite{Burrows:2017wqn}. Further details of the calculations
of the effective potential and the NA scattering 
are described in Ref.~\cite{Burrows:2018ggt}.

First, we want to illustrate
the difference in employing $U$ or ${\hat U}$ as effective NA potential in
Fig.~\ref{fig1} for proton scattering from $^4$He and $^{16}$O at 200~MeV projectile laboratory kinetic
energy. The figure shows that taking into account the effect of the
operator $Q$ by solving
Eq.~(\ref{eq:4}) to obtain $U$ is clearly visible for the light nucleus
$^4$He, while very small for a heavier nucleus like $^{16}$O. 

As examples for elastic proton scattering based on an {\it ab initio}
effective potential we show in Fig.~\ref{fig2} the angular distributions
(divided by the Rutherford amplitude) for $^4$He and $^{16}$O for
energies between 100 and 200~MeV laboratory kinetic energy. We find them
in very good agreement with the data in the angle and momentum transfer
regime where the first order term of the full-folding effective
potential should be valid. We also want to point out that the first
order term in the multiple scattering expansion does not explicitly
contain any 3NF contributions, thus the choice of the NNLO$_{\rm{opt}}$
interaction works well with the theoretical content of the effective
potential.  Further in the future with different interactions will have
to shed more light on the effect of including 3NFs in the one-body
density for the first-order effective potential.
 
\vspace{2mm}
\noindent
\footnotesize{
{\bf Acknowledgement:} Partial support for this work is given by the
U.S. DoE under DE-FG02-93ER40756, DE-SC0018223, DE-AC02-05CH11231, and
the U.S. NSF under OIA-1738287, ACI-1713690,  OCI-0725070, and
ACI-1238993.
}
%
%
\vspace{-3mm}
\bibliographystyle{spphys}
\bibliography{clusterpot,denspot,ncsm}

\end{document}